\documentstyle[12pt,twoside,psfig,epsfig]{article}
\begin{document}
\begin{center}
{\Large\bf Spherically symmetric scalar field collapse}
\\[15mm]
Koyel Ganguly$^{*}$ {\footnote{E-mail: koyel\_g\_m@yahoo.co.in}} and
            Narayan Banerjee$^{\dagger}${\footnote {E-mail: narayan@iiserkol.ac.in}}\\
$^{*}$Relativity and Cosmology Research Centre, Department of Physics,
        Jadavpur University, Kolkata - 700032, India.\\
 $^{*}$St. Xavier's College, 30 Mother Teresa Sarani, Kolkata - 700016, India.\\
$^{\dagger}$ IISER - Kolkata, Mohanpur Campus, P.O. BCKV Main Office, District Nadia,\\ West Bengal 741252, India.
\end{center}
\date{}
\vspace{0.5cm}
\vspace{0.5cm}
\pagestyle{myheadings}
\newcommand{\be}{\begin{equation}}
\newcommand{\ee}{\end{equation}}
\newcommand{\bea}{\begin{eqnarray}}
\newcommand{\eea}{\end{eqnarray}}

\begin{abstract}
It is shown that a scalar field, minimally coupled to gravity may
have collapsing modes even when the energy condition is violated,
that is, for $(\rho+3p)<0$. This result may be useful in the
investigation of the possible clustering of dark energy. All the 
examples dealt with have apparent horizons which form before
the  formation of the singularity. The singularities formed are
shell focusing in nature. The density of the scalar field
distribution is seen to diverge at singularity. The Ricci
scalar also diverges at the singularity. The interior spherically
symmetric metric is matched with exterior Vaidya metric at the
hypersurface and the appropriate junction conditions are obtained.
\end{abstract}

\section{Introduction}
\par The ultimate fate of a collapsing matter distribution remains a subject of fervent interest for many years. It is generally believed that a collapsing matter distribution, if it satisfies the energy condition $(\rho+3p)>0$, will crush to a singularity. There is however no general indication whether an event horizon would form to shield this singular state of matter from a distant observer. In fact, there are ample theoretical examples where the singularities could be naked to an external observer either for a finite period of time or even for ever. For a comprehensive review we refer to the monograph by Joshi\cite{1}. A lot of work therein or that carried out later normally considers the collapse of a perfect fluid. As the final stages of collapse might involve dissipative processes as well, some later investigations also consider viscous effects (see \cite{2} and references therein).
\par Investigations on collapsing gravitational systems with a scalar field revealed that the non linearity of Einstien equations could lead to critical phenomena close to the threshold of black hole formations\cite{3}. Naturally this led to an arena of great interest(for recent reviews we refer to
\cite{4}). Furthermore, the scalar field collapse could also lead to the formation of naked singularities. This was demonstrated by analytical investigations by Christodoulou\cite{5}, Goswami and Joshi\cite{8}, Giambo\cite{12} and also by numerical work of Choptuik\cite{3}, Brady\cite{6}, Gundlach\cite{7}, and others.
\par In a series of papers Christodoulou\cite{christ2, christ3, christ4} pioneered analytical studies in gravitational collapse of a spherically symmetric configuration of massless scalar field models. Detailed investigations on such models have been carried out later\cite{3, 6, goldwirth}. Collapsing spherically symmetric massive scalar field models have also appeared in the literature\cite {12, gonclaves1, gonclaves2, joshi}. Gonclaves\cite{gonclaves1} investigated the black hole formation with a massive scalar field in an Einstein-de Sitter universe. Gonclaves and Moss\cite{gonclaves2} also showed that for a range of parameters, the collapse of a spherically symmetric scalar field can be formally treated as a collapsing dust ball. This is similar to the case of $\gamma=\frac{1}{2}$ in the present work (section 3.3). Giambo\cite{12} showed that a massive scalar field, if collapses completely, can give rise to a naked singularity. Goswami and Joshi\cite{8} constructed a
 class of solutions with a massive scalar field which could lead to a naked singularity.
\par The very recent interest in scalar field collapse stems from a cosmological requirement. It has already become a folklore that the present Universe is
undergoing an accelerated phase of expansion. Scalar fields endowed with a potential are amongst the most favoured candidates as the ``dark energy''- the agent driving the cosmic acceleration\cite{9}. Although the general speculation is that the dark energy does not cluster at any scale below the Hubble scale, there is not much evidence for this. The collapse dynamics of a minimally coupled scalar field with a negative effective pressure can indeed help us to have a better understanding of the scenario.
\par The objective of the present investigation is to explore the collapsing modes of a simple spherically symmetric minimally coupled scalar field model.
The scalar field $\phi$ is assumed to depend only on time, so this is an analogue of Oppenheimer - Snyder model of the collapse of a spatially homogeneous fluid\cite{10}.
\par Section $2$ describes the scalar field distribution and obtains the general first integrals in situations where the effective pressure $p_{\phi}$ and energy density $\rho_{\phi}$ of the scalar field are related as $p_{\phi}=w\rho_{\phi}$, $w$ being a constant. Section $3$ discusses special cases of the collapsing models. In section $4$ the collapsing spacetime is matched at the boundary with an external metric and in section $5$ some concluding remarks are made.
\section{Spherically symmetric time dependent distribution of a scalar field}
The relevant action for a scalar field minimally coupled to gravity is given by
\be\label{action}
{\mathcal{A}}=\int\sqrt{-g}d^4 x[R +\frac{1}{2}{{\phi}^{,\mu}{\phi}_{,\mu}}-V(\phi)].
\ee
We take the spherically symmetric metric as
\be\label{metric}
ds^2=dt^2-A^2(r,t)dr^2-B^2(r,t)d{\Omega}^2,
\ee
where A and B are functions of the radial coordinate `$r$' and time `$t$'. A minimally coupled scalar field $\phi$ endowed with a potential $V=V(\phi)$
is considered. For the sake of simplicity, the scalar field is taken to be spatially homogeneous and evolving with time alone. The non-trivial field equations become
\be\label{fe1}
-\frac{2{B}^{\prime\prime}}{A^2B}+\frac{2{A}^{\prime}{B}^{\prime}}{A^3B}-\frac{{{B}^{\prime}}^2}{A^2B^2}+2\frac{\dot{A}\dot{B}}{AB}+\frac{{\dot{B}}^2}{B^2}+\frac{1}{B^2}
=\frac{1}{2}{\dot{\phi}}^2+V(\phi),
\ee
\be\label{fe2}
\frac{{{B}^{\prime}}^2}{A^2B^2}-2\frac{\ddot{B}}{B}-\frac{{\dot{B}}^2}{B^2}-\frac{1}{B^2}=\frac{1}{2}{\dot{\phi}}^2-V(\phi),
\ee
\be\label{fe3}
\frac{{B}^{\prime\prime}}{A^2B}-\frac{{A}^{\prime}{B}^{\prime}}{A^3B}-\frac{\ddot{A}}{A}-\frac{\ddot{B}}{B}-\frac{\dot{A}\dot{B}}{AB}=
\frac{1}{2}{\dot{\phi}}^2-V(\phi),
\ee
\be\label{fe4}
\frac{B^{\prime}}{B}\frac{\dot{A}}{A}-\frac{{\dot{B}}^{\prime}}{B}=0,
\ee
where $\phi=\phi(t)$ is the scalar field, $V(\phi)$ is the scalar potential. The equations are written in units where $8{\pi}G=1$. An overhead dot and prime indicates differentiation w.r.t $`t'$ and $`r'$ respectively. As already mentioned, the contribution from matter (the scalar field in this case) is spatially homogeneous like the homogeneous fluid collapse discussed by Oppenheimer and Snyder\cite{10}. Equation (\ref{fe4}) readily integrates to yield
\be\label{bprime}
B^{\prime}=\lambda(r)A,
\ee
where $\lambda(r)$ is an arbitrary function of `$r$'. In what follows $\lambda(r)$ is taken to be constant for the sake of simplicity. This indeed is a special case, but these would lead to some tractable solutions so that the possibility of collapse can be investigated. However, if $A(r,t)$ is separable as a product of functions of `$r$' and `$t$', the assumption (\ref{bprime}) is hardly restrictive as $\lambda(r)$ in that case can be absorbed by the scaling of the radial co-ordinate. Anyway, the physical motivation is that in this case the collapsing model will eventually become an FRW one. If $A$ and $B$ are now written as $A^2=e^{\alpha}$ and $B^2=e^{\beta}$, the field equations take a much simpler form as
\be\label{sfe1}
\frac{{\dot{\beta}}^2}{4}+\frac{\dot{\alpha}\dot{\beta}}{2}=\frac{1}{2}{\dot{\phi}}^2+V(\phi),
\ee
\be\label{sfe2}
\ddot{\beta}+\frac{3}{4}{\dot{\beta}}^2=-\frac{1}{2}{\dot{\phi}}^2+V(\phi),
\ee
\be\label{sfe3}
\frac{\ddot{\alpha}}{2}+\frac{{\dot{\alpha}}^2}{4}+\frac{\ddot{\beta}}{2}+\frac{{\dot{\beta}}^2}{4}+\frac{\dot{\alpha}\dot{\beta}}{4}=-\frac{1}{2}{\dot{\phi}}^2+V(\phi),
\ee
where equation (\ref{bprime}) has been used, which now reads as
\be\label{ealpha}
e^{\alpha}=\frac{{\beta^{\prime}}^2}{4}e^{\beta}.
\ee
The wave equation for the scalar field is given by
\be\label{waveeq}
{{\phi}^{,\mu}}_{;\mu}+V_{,\phi}=0.
\ee
In what follows, the potential $V(\phi)$ will be assumed to be proportional to ${\dot{\phi}}^2$ ,
\be\label{vchoice}
V(\phi)=\gamma{\dot{\phi}}^2,
\ee
where $\gamma$ is a constant. Indeed this choice does not have any physical requirement, but along with a major simplification for the equation system the ansatz provides the fourth equation which we require to solve the system of equation. We have only three independent equations (\ref{sfe1})-(\ref{sfe3}) to solve for four unknowns $\alpha$, $\beta$, $\phi$ and $V$.  As $g_{00}=1$ and $\phi=\phi(t)$, ${\dot{\phi}}^2$ is actually ${\phi}^{,\mu}{\phi}_{,\mu}$ which is indeed a scalar and equation (\ref{vchoice}) is thus consistent. One can see that $\gamma>\frac{1}{2}$ would lead to the case where the effective pressure is negative.
\par However, it deserves mention at this stage that the field in this case is not a k-essence field where the scalar field Lagrangian is given by $L(X)$, where $X=(\nabla\phi)^2$. Actually the model does have a potential and the present work deals with those which are proportional to the kinetic part. We shall see later that an exponential potential is in fact consistent with the ansatz(\ref{vchoice}) in the present model.
\par The equation of state parameter, $w=\frac{p_{,\phi}}{\rho_{,\phi}}=\frac{\frac{1}{2}{\dot{\phi}}^2-V(\phi)}{\frac{1}{2}{\dot{\phi}}^2+V(\phi)}$ will be a constant given by $w=\frac{1-2\gamma}{1+2\gamma}$. For $\gamma\geq 1$, the equation of state parameter $\omega$ is more negative than $-\frac{1}{3}$ and the scalar field acts as a dark energy. For any finite positive value of $\gamma$, $w$ remains greater than $-1$ and hence the scalar field is in fact a quintessence field and cannot give a phantom field with $w<-1$. If $\gamma=0$, that is, for the case without potential, the value of $w$ becomes 1 and the field is then formally similar to a stiff fluid. For $\gamma = \frac{1}{2}$, one has an effectively zero pressure, and the scalar field is formally equivalent to a pressureless dust.
\par With the help of equation (\ref{vchoice}), the wave equation (\ref{waveeq}) yields a first integral as,
\be\label{phidot}
{\dot{\phi}}^{(2\gamma+1)}=\frac{1}{e^{(\frac{\alpha}{2}+\beta)}}g(r),
\ee
where $g(r)$ is an arbitrary function of $r$ to be determined from initial conditions.
As $\phi$ is assumed to be spatially homogeneous, this equation indicates,
\be\label{exp alphabeta}
e^{(\frac{\alpha}{2}+\beta)}=g(r)f(t),
\ee
where $f(t)$, a function of time, and has to be determined from the field equations. This gives the form of equation (\ref{phidot}) as,
\be\label{phidot1}
\dot{\phi}=f(t)^{-\frac{1}{1+2\gamma}}.
\ee
Using (\ref{exp alphabeta}) in a combination of field equations (\ref{sfe1})-(\ref{sfe3}) one obtains
\be\label{fddot}
\ddot{f}=3\gamma{[f(t)]}^\frac{2\gamma-1}{2\gamma+1},
\ee
which readily yields a first integral as
\be\label{fdot}
{\dot{f}}^2=\frac{3}{2}(2\gamma+1)f^{\frac{4\gamma}{2\gamma+1}}+C_{1},
\ee
$C_{1}$ being an arbitrary constant. In order to satisfy the field equations, one can verify that $C_{1}=0$. Choosing $g(r)$ as $r^2$ the solution of
$e^{\alpha}$ and $e^{\beta}$ are given as,
\be\label{ealpha1}
e^{\alpha}=(f(t))^{2/3},
\ee
and
\be\label{ebeta}
e^{\beta}=(f(t))^{2/3}r^2.
\ee
Equation (\ref{fdot}) can be written as (with $C_{1}=0$),
\be\label{fdotc1zero}
\dot{f}=-\sqrt{\frac{3}{2}(2\gamma+1)}f^{\frac{2\gamma}{(2\gamma+1)}}.
\ee
The negative root is chosen in order to consider a collapsing scenario. Equation (\ref{fdotc1zero}) readily integrates to yield
\be\label{ft}
f={\left(\frac{t_{0}}{2\gamma+1}-\sqrt{\frac{3}{2(2\gamma+1)}}t\right)}^{(2\gamma+1)},
\ee
where $t_{0}$ is another constant of integration.
\par The solution of the scalar field $\phi$ as some function of $t$ is given by (using equations (\ref{phidot}) and (\ref{exp alphabeta})),
\be\label{phi}
\phi=-\sqrt{\frac{2(2\gamma+1)}{3}}ln{\left(\frac{t_0}{2\gamma+1}-\sqrt{\frac{3}{2(2\gamma+1)}}t\right)}.
\ee
\par As the complete set of solutions is now in hand, it is not difficult to check which form of potential is consistent with the ansatz(\ref{vchoice}). It is easy to see from equation (\ref{phi}) that
\be\label{pot1}
{\dot{\phi}}^2
= {\left(\frac{t_0}{2\gamma+1}-\sqrt{\frac{3}{2(2\gamma+1)}}t\right)}^{-2}
=exp(\frac{2\phi}{\sqrt{\frac{2(2\gamma+1)}{3}}}).
\ee
In conjunction with equation (\ref{vchoice}), one can now write,
\be\label{pot2}
V(\phi)={\gamma}exp{\left(\frac{2\phi}{\sqrt{\frac{2(2\gamma+1)}{3}}}\right)}.
\ee
\section{Collapsing models}
\subsection{$\gamma<1/2$}
For values of $\gamma<1/2$, the matter content satisfies the energy condition ($\rho +3p>0$). Choosing $\gamma=1/4$, one has the solution of $f(t)$ as
\be\label{ft1}
f(t)={\left(\frac{2}{3}t_{0}-t\right)}^{3/2}
\ee
Now $e^{\frac{\alpha}{2}+\beta}$, which according to equation (\ref{exp alphabeta}) is  given by
$g(r)f(t)$, determines the proper volume. The above equation clearly shows that $f(t)$ indeed has a non zero value for all $t$ less that $\frac{2t_0}{3}$ but becomes zero when time $t=\frac{2}{3}t_0$. So for this value of $t$ (designated as $t_{s}$), one has a singularity of a zero volume. Choosing a value of $t_0=3$, one gets $t_{s}=2$. The plot of $f(t)$ vs $t$ in Figure $1$ clearly reveals the collapsing scenario.
\par The condition for the apparent horizon is
\be\label{ah} g^{\alpha\beta}R_{,\alpha}R_{,\beta}=0, \ee where
$R$ is the proper radius of the two sphere. In the present case
$R=B=e^{{\beta}/2}$ and equation (\ref{ah}) reads like
\be\label{ah1} g^{00}{\dot{B}}^2+g^{11}{B^{\prime}}^2=0, \ee that
is \be\label{ah2} {\dot{\beta}}^2-e^{-\alpha}{\beta^{\prime}}^2=0,
\ee which in view of (\ref{ealpha}) yields
\be\label{ebeta1}
e^{\beta}{\dot{\beta}}^2=4. \ee
 The time of formation of the apparent horizon , $t_{ah}$, has been calculated for a fixed value of $r=1$ in all the cases discussed here. In this case ($\gamma=1/4$ and $t_0$ is chosen as 3), $t_{ah}=1.967$. Hence the singularity formed is shielded by the horizon.
\begin{figure}[!h]
\centerline{\psfig{figure=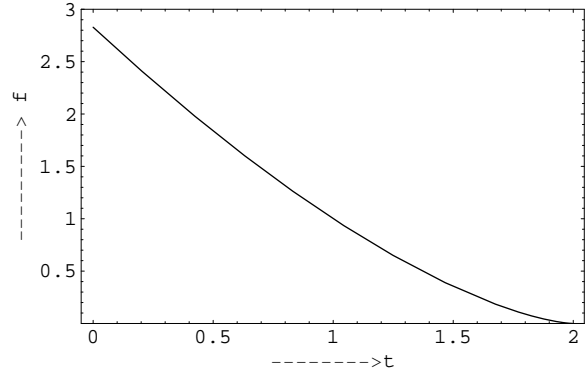,height=50mm,width=80mm}}
\caption{\normalsize{Plot of $f(t)$ vs. $t$ for $\gamma = 1/4$ putting $t_{0}=3$.}}
\label{fig(i)}
\end{figure}

\subsection{$\gamma>\frac{1}{2}$}

\par Choosing a value of  $\gamma>\frac{1}{2}$ makes $\rho +3p<0$, that is the energy condition is violated and the scalar field is apt to act as a dark energy. With a choice of $\gamma=3/2$(as an example), the solution of $f(t)$ turns out to be,
\be\label{ft2}
f(t)={\left(\frac{t_0}{4}-\sqrt{\frac{3}{8}t}\right)}^4.
\ee
It is evident from the above expression that $f(t)$ becomes zero when $t=\frac{t_0}{\sqrt{6}}$. Hence one indeed has a
singularity of a zero proper volume
 $(f=0)$ at a finite future. Choosing $t_0=4$ one gets $t_{s}=1.632$ and $t_{ah}=0.636$. Figure $2$ shows the plot of $f(t)$
 against time for choice of $t_0=4$. It is observed that initially there is a non zero proper volume and then the
 singularity of zero volume is attained at a finite time. A change in initial condition changes the time at which
 singularity is attained keeping the nature of the plots unchanged. The time at which apparent horizon is formed
 also changes with the change of initial condition. For the choice of $t_0$=4, $t_{ah}<t_{s}$. Hence the singularity
 formed  is covered by the horizon from an external observer.
 \begin{figure}[!h]
\centerline{\psfig{figure=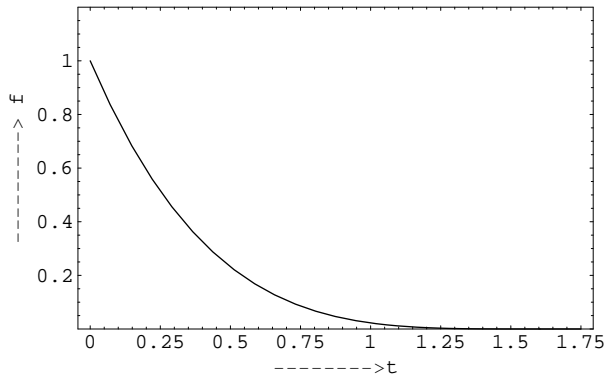,height=50mm,width=80mm}}
\caption{\normalsize{Plot of $f(t)$ vs. $t$ for $\gamma = 3/2$
putting $t_{0}=4$.}} \label{fig(ii)}
\end{figure}
\begin{figure}[!h]
\centerline{\psfig{figure=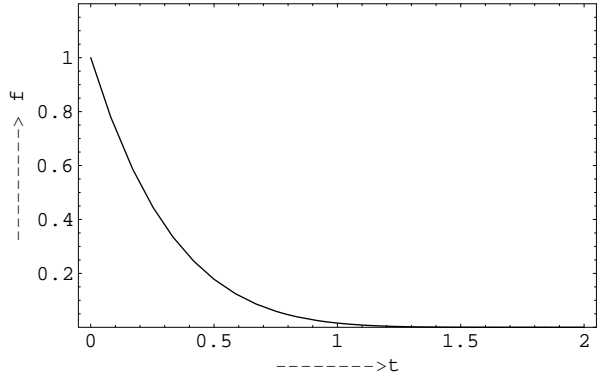,height=50mm,width=80mm}}
\caption{\normalsize{Plot of $f(t)$ vs. $t$ for $\gamma = 5/2$
putting $t_{0}=6$.}} \label{fig(iii)}
\end{figure}

\par For some other choice of  $\gamma>1/2$, for example $\gamma=5/2$ and choosing $t_0=6$, $f(t)$ vs $t$ is plotted
as shown in Figure $3$. The plot shows that in this case also  $f(t)$ becomes zero at some finite time $(t_{s}=2)$.
 The value of $t_{ah}$ is $0.592$. In all the cases $t_{ah}<t_{s}$, thus the singularities formed are hidden behind
 the horizon. 
 \par The energy density and the effective pressure of the scalar field are given by $\rho_{\phi}=(\frac{1}{2}+\gamma){f(t)}^{-\frac{2}{1+2\gamma}}$ and $p_{\phi}=(\frac{1}{2}-\gamma){f(t)}^{-\frac{2}{1+2\gamma}}$ respectively. From the expressions for $f(t)$ it is clear that for any positive value of $\gamma$ whether it is greater than or less than 1/2, the density and pressure of the scalar field would diverge at the time at which the proper volume becomes zero. The Ricci scalar $R$ also diverges at the singularity as is evident from its expression given below,
 \begin{equation}\label{ricci2}
 R=\frac{-\ddot{f}}{f}+\frac{1}{9}\frac{{\dot{f}}^2}{f^2}+\frac{36f^{-2/3}}{r^2}.
 \end{equation}

\subsection{$\gamma=1/2$}
\par For $\gamma=1/2$, the model formally resembles a perfect fluid
distribution with zero pressure. From equation (\ref{ft}), one has
the solution of $f(t)$ as
\be\label{ft0}
f(t)={\frac{(t_{0}-\sqrt{3}t)}{2}}^2.
\ee
At time
$t=\frac{t_0}{\sqrt{3}}$ the proper volume becomes zero. Choosing
$t_0=2$, one gets $t_{s}=1.155$. Fig $4$ shows the plot of $f(t)$
vs $t$ for $\gamma=1/2$.
\begin{figure}[!h]
\centerline{\psfig{figure=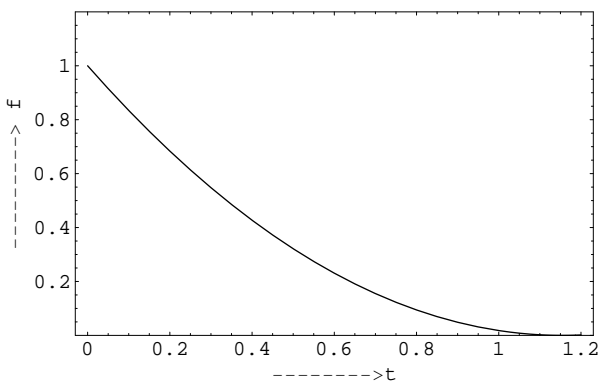,height=50mm,width=80mm}}
\caption{\normalsize{Plot of $f(t)$ vs. $t$ for $\gamma = 1/2$ putting $t_{0}=2$.}}
\label{fig(iv)}
\end{figure}
The time of apparent horizon formation in this case is $t_{ah}=1.071$ which is once again less than the value of $t_{s}$.
\par Singularities formed in collapse can be shell focussing or shell crossing in nature\cite{yodzis, waugh}.
For a spherically symmetric collapse the shell focusing
singularity occurs at $g_{\theta\theta}=0$ and the shell crossing
singularity occurs when
$\frac{dg_{\theta\theta}}{dr}=0,g_{\theta\theta}\neq 0$ and $r>0$.
It is evident from (\ref{metric}) and (\ref{ebeta}) that
$g_{\theta\theta}=r^4{f(t)}^{2/3}$. At singularity since $f(t)=0$,
therefore $g_{\theta\theta}=0$ at singularity, clearly indicating
that the singularities formed in all the cases are shell focusing
in nature.

\section{Matching with the exterior metric}
\par The interior metric now has to be matched with an exterior metric at the boundary $\Sigma$. Following ref \cite{8},  we match the spherical ball of collapsing scalar field to a generalized Vaidya metric . The metric just inside $\Sigma$ is,
\begin{equation}\label{interior}
d{s_-}^2=dt^2-e^{\alpha}dr^2-e^{\beta}d{\Omega}^2,
\end{equation}
and the metric in the exterior of $\Sigma$ is given by
\begin{equation}\label{exterior}
d{s_+}^2=(1-\frac{2M(r_v,v)}{r_v})dv^2+2dvdr_v-{r_v}^2d{\Omega}^2.
\end{equation}
Matching the first fundamental form on the hypersurface we get
\begin{equation}\label{cond1}
{\frac{dv}{dt}}_{\Sigma}=\frac{1}{\sqrt{1-\frac{2M(r_v,v)}{r_v}+\frac{2dr_v}{dv}}}
\end{equation}
and
\begin{equation}\label{cond2}
r_v=r{f(t)}^{1/3}
\end{equation}
Matching the second fundamental form yields,
\begin{equation}\label{cond3}
e^{\beta/2}=r_v\left(\frac{1-\frac{2M(r_v,v)}{r_v}+\frac{2dr_v}{dv}}{\sqrt{1-\frac{2M(r_v,v)}{r_v}+\frac{2r_v}{dv}}}\right)
\end{equation}
Using equations \ref{cond1}, \ref{cond2} and \ref{cond3} one can write
\begin{equation}\label{dvdt2}
\frac{dv}{dt}=\frac{1-\frac{rf^{-2/3}}{3}}{1-\frac{2Mf^{-1/3}}{r}}.
\end{equation}
From equation \ref{cond3} one obtains
\begin{equation}\label{M}
M=\frac{r^{-1}f^{-1/3}+\frac{rf^{-5/3}}{9}+\sqrt{\frac{1}{r^{2}f^{2/3}}+\frac{r^2}{81f^{10/3}}-\frac{2}{9f^{2}}}}{\frac{4}{r^2f^{2/3}}}.
\end{equation}
Matching the second fundamental form we can also write the derivative of $M(v,r_v)$ as
\begin{equation}\label{dM}
M{(r_v,v)}_{,r_v}=\frac{M}{rf^{1/3}}-\frac{2r^2}{9f^{4/3}}.
\end{equation}
Equations \ref{cond2}, \ref{dvdt2}, \ref{M} and \ref{dM} completely specify the matching conditions at the boundary of the
collapsing scalar field.

\section{Discussions}
The present work investigates a simple spherically symmetric spacetime with a scalar field. The scalar field is
self-interacting, i.e., it has a potential. The field is chosen as spatially homogeneous, so it represents
 the scalar field analogue of Oppenheimer-Snyder collapse for a perfect fluid \cite{10}. At least for the special case
 where the potential could be written as proportional to the kinetic part of $\phi$, it could be shown that the model
 possesses a collapsing solution.
\par Most of the models on massive scalar field collapse, available in the literature \cite{12, gonclaves1, gonclaves2},
demonstrate the formation of either a naked singularity or a black hole are subject to the initial conditions.
They generally work it out for energy condition $(\rho+3p)>0$. In our work we demonstrate the collapse of a scalar
field endowed with a potential for both the energy conditions $(\rho+3p)>0$ and $(\rho+3p)<0$. The value of the equation of state parameter $w$, determined by the value of $\gamma$, dictates whether the energy condition is satisfied or not.When $\gamma$ is less that $-1$, the energy condition is violated.  For $(\rho+3p)>0$,
like many other models\cite{8, 12, joshi}, we get singularities at a finite time and the possibility of the formation of an
apparent horizon ahead of the collapse to the singularity.
 But the intriguing feature of the present work is that we obtain singularities at a finite time even for the condition
 $(\rho+3p)<0$ wherein the effective pressure is negative.
 So this is an investigation of the collapse scenario of dark energy itself rather than collapse of a fluid against
  a dark energy background\cite{TS}. This is an indication towards the possibility of the clustering of dark energy,
 where $(\rho + 3p) < 0$.
\par All the examples discussed in the present work allow the formation of an apparent horizon well
 before the distribution crushes to the singularity of zero volume. So the singularities are not naked. The interior metric has been matched with the exterior
Vaidya metric and the appropriate matching conditions are obtained.
\par In all the cases (for different values of $\gamma$ chosen) it is observed that the energy  density as well
as the effective pressure for the scalar field diverge at the singularity. It has been checked that the scalar
curvature $R = R^{\mu}_{\mu}$ also diverges as the proper volume becomes zero. The scalar field $\phi$ also is singular
 at $t_{s}$.  Equations (\ref{ft1}) and (\ref{ft2}) have a deceiving feature, as if although the volume is zero at $t = t_{s}$,
 it becomes quite well behaved beyond $t_{s}$. A look at equation (\ref{phi}) reveals that actually beyond $t_{s}$
 the scalar field itself becomes undefined as the argument of logarithm becomes negative. So one indeed has a
 singularity at a finite future, beyond which the model cannot be discussed any further.
\par The present work deals with a particular potential and does not represent a general kind of potential.
Although we started with an assumption that the potential is proportional to the kinetic part, actually the
model is that of an exponential potential (\ref{pot2}).
\par

\vskip .5in

{\large\bf Acknowledgement:}\\

\par The authors thank the BRNS (DAE) for financial support.\\

\end{document}